\documentclass[twocolumn,prd,amsmath,amssymb]{revtex4}
\input psfig.sty 
\def \be {\begin{equation}} 

\def \ee {\end{equation}} 
\def \bea {\begin{eqnarray}} 
\def \eea {\end{eqnarray}}

\usepackage{graphicx}
\usepackage{dcolumn}
\usepackage{bm}
\usepackage{epsfig}

\begin{document} 

\title{Dodging the dark matter degeneracy while determining the dynamics of dark energy}

\author{Vinicius C. Busti$^{1,2}$} \email{vinicius.busti@iag.usp.br} 

\author{Chris Clarkson$^{1}$} \email{chris.clarkson@uct.ac.za} 

\vskip 0.5cm \affiliation{$^{1}$ Astrophysics, Cosmology \& Gravity Centre, and Department of Mathematics and Applied Mathematics, 
University of Cape Town, Rondebosch 7701, Cape Town, South Africa \\
$^{2}$ Departamento de F\'{i}sica Matem\'{a}tica, Instituto de F\'{i}sica, Universidade de S\~{a}o Paulo,  \\
CEP 05508-090, S\~{a}o Paulo - SP, Brazil } 


\begin{abstract} 
\noindent 
One of the key issues in cosmology is to establish the nature of dark energy, and to determine whether the equation of state evolves with time. When estimating this from distance 
measurements there is a degeneracy with the matter density. We show that there exists a simple function of the dark energy equation of state and its first derivative which is 
independent of this degeneracy at all redshifts, and so is a much more robust determinant of the evolution of dark energy than just its derivative. We show that this function can be well 
determined at low redshift from supernovae using Gaussian Processes, and that this method is far superior to a variety of parameterisations which are also subject to priors on 
the matter density. This shows that parametrised models give very biased constraints on the evolution of dark energy.

\end{abstract} 

\maketitle 

\section{Introduction.}

Since the discovery in 1998 that the Universe is undergoing a period of accelerated expansion \cite{riess1998,perlmutter1999},
considerable work has been done trying to understand what it is driving the phenomenon. The standard explanation beyond a simple cosmological constant relies on introducing a new cosmic fluid, dubbed dark energy, which possesses negative pressure in order to accelerate the expansion.
Dark energy is parametrised via an equation of state $w(z)=\frac{p_{DE}}{\rho_{DE}}$, where $p_{DE}$ is its pressure, $\rho_{DE}$ is its energy density and $w(z)$ stands for a 
possible dependence of the equation of state with the redshift $z$. The standard case is given by $w(z)=-1$, which represents the cosmological constant $\Lambda$, where together 
with cold dark matter (CDM) form the cosmic concordance model $\Lambda$CDM.

Extensive observational efforts have been carried out in order to falsify $w(z)=-1$ through several observables, as type Ia supernovae (SNe Ia),
temperature anisotropies of the cosmic microwave background (CMB) and baryon acoustic oscillations (BAOs). As there is no compelling model to explain the acceleration, it is often simplest to use simple parameterisations for $w$, where a dependence with the redshift can be accommodated by adding a new parameter.
On the other hand, an interesting and independent way to test for deviations of the $\Lambda$CDM model is to use non-parametric methods, which are suitable for describing 
how $w(z)$ evolves not restricted by simple parametric models. Examples considered in the literature include the principal component analysis method (PCAs) \cite{pca}, 
gaussian processes (GPs) \cite{gp1,gp2}
and local regression smoothing \cite{lrs}. 

Reconstructions have been limited by degeneracies involving the matter density parameter $\Omega_m$ and the curvature 
density parameter $\Omega_k$. From one side, $\Omega_k$ can be inferred due its geometrical structure \cite{omega_k}, whereas for $\Omega_m$ there is a so-called dark degeneracy
\cite{kunz2009} whereby dark energy and dark matter are hard to separate. 

Here we show that there is a simple function of $w(z)$ and its first derivative which is completely independent of $\Omega_m$. 
This function can be used to look for a time evolution of the dark energy equation of state without any dark matter degeneracy.
We use data from SNIa and $H(z)$ measurements 
from cosmic chronometers and BAOs to constrain this new function, along with the evolution of $w$. We compare parametric reconstructions to a GP reconstruction and  show that GPs provide much stronger constraints than parameterisations at low redshifts and are broadly consistent with $\Lambda$CDM.

\section{The evolution of dark energy and the distance}

The dimensionless comoving distance $D(z)$ can be written as a function of the luminosity distance $d_L(z)$ as
\begin{equation}
D(z)=(H_0/c)(1+z)^{-1}d_L(z),
\end{equation}
where $H_0$ is the Hubble constant. The effective dark energy equation of state is
\begin{equation}\label{djskbcskcb}
w(z) = \frac{2(1+z)D'' + 3D'}{3\Omega_m(1+z)^3D'^3 - 3D'},
\end{equation}
where a prime denotes differentiation with respect to the redshift. Throughout the paper we assume $\Omega_k=0$, although we note a degeneracy between $w$ and $\Omega_k$ can be 
important at high redshifts
\cite{degeneracy_w_om_k}. Given distance-redshift data, we can determine $w$ from this relation. However, the value of $\Omega_m$ must be provided {\it a priori} from independent 
methods in order to test for deviations of $w=-1$ \cite{om-w}. Similarly, one can write the effective adiabatic sound speed for dark energy as
\begin{eqnarray}
c_s^2(z) & = & w(z) + \frac{1}{3}(1+z)\frac{w^{\prime}}{1+w(z)},\\
 & = &
\frac{2}{3} \frac{(1+z)D'D''' - 2D'D'' - 3(1+z)D''^2}{D'
  \left[2D'' + 3\Omega_m (1+z)^2 D'^3 \right]}\,.         
\end{eqnarray}
Combining with $w(z)$ we define
\begin{eqnarray}
\label{eqF}
F(z) &=& \frac{1+w(z)}{w(z)} c_s^2(z) \\
&=& \frac{1}{3} (1+z) \frac{w'(z)}{w(z)} + 1 + w(z) \nonumber\\
&=& \frac{2}{3} \frac{(1+z)\left[(1+z)D'D''' - 2D'D'' -
    3(1+z)D''^2\right]}{D'\left[2(1+z)D'' + 3D'\right]}.\nonumber    
\end{eqnarray}
Note that $F(z)$ is completely independent of $\Omega_m$ and is a function of the observable quantity $D$ and its derivatives. Model-independent constraints on $F$~-- and consequently $w$~-- can be 
derived through non-parametric methods.  Any deviation from $F=0$ would rule out the flat $\Lambda$CDM model, and so forms a useful null test of concordance cosmology~\cite{foot} (see \cite{sahni2014} for another null test independent of $\Omega_m$). Another way to think of the function $F$ is to say that in the possible function space of $w(z)$, $F$ selects the subspace which is independent of $\Omega_m$, while giving a measure of the first derivative of $w$.

\subsection{Constraining $w$ and $F$}

We shall now constrain $F$ in 2 different ways. Given SNe Ia and $H(z)$ data, we need to find $D(z)$ and its first 3 derivatives. First we will use Gaussian Processes (GPs) which are a robust 
smoothing technique which allows for differentiation of data~-- this technique is summarized in the Appendix. We will use the third line in Eq. \ref{eqF} to obtain constraints on $F$ and Eq. \ref{djskbcskcb} with a suitable prior on $\Omega_m$ to derive constraints for $w$.  Then we will compare these results with two parametric models. 
We choose a standard one and a new one which produces similar constraints on $w$ but rather different constraints on the evolution.

\begin{itemize}

\item A standard series expansion in $1-a$: $w(z) = w_0 + w_a \frac{z}{1+z}$, generally called the CPL (Chevalier-Polarski-Linder) parameterisation
\cite{cpl}. 

\item Constant $F$: $w(z) = \frac{w_0 C}{(w_0 + C)(1+z)^{3C} - w_0}$.
This model is constructed such that $F(z) = 1-C$ is constant.
Note that $C=1$ gives $F=0$, which means that although $F \neq 0$ rules out flat $\Lambda$CDM, $F=0$ is also compatible with
different models.

\end{itemize}

Although the second of these may look rather unusual, these are actually rather similar in terms of their basic assumptions because the CPL parameterisation is constructed so that $w_{,a}$ is constant.

\subsection{Data}

For both the non-parametric and parametric analyses we use 580 luminosity distance measurements from SNe Ia \cite{union2.1}, where we included
all the systematic errors, and 26 $H(z)$ measurements from cosmic chronometers \cite{hz_cc} and BAOs \cite{hz_bao}.
For the Hubble constant we adopt $H_0=70.4\pm2.5$ km s$^{-1}$ Mpc$^{-1}$ \cite{Komatsu:2010fb}, which is in agreement with the revisited
local value based on the NGC 4258 maser distance obtained by Efstathiou \cite{efs}: $H_0=70.6\pm3.3$ km s$^{-1}$ Mpc$^{-1}$ .

As we assume spatial flatness ($\Omega_k=0$), the $H(z)$ measurements
can be directly used to obtain the first derivative of $D$: $D'(z) = \frac{H_0}{H(z)}$,
which allows us to incorporate the derivative of $D$ as data in our GPs analysis.

For the parametric analyses, we obtain the posterior probabilities (see e.g. \cite{ratra}) using emcee \cite{emcee}, an affine-invariant ensemble sampler for Markov Chain 
Monte Carlo (MCMC) \cite{affine}. For example, in the CPL case we use the posteriors of $w_0-w_a$ to derive constraints to $w(z)$ sampling from these parameters, 
the same for $F$ using the second line of Eq. \ref{eqF}.

An important issue in our analysis is that distance moduli of SNe Ia were used already fitted for some nuisance parameters in
a given cosmological model \cite{union2.1}. Therefore, a more model-independent procedure would be to introduce these nuisance parameters in our
GP regression. On the other hand, such approximation was also taken in our parametric analyses, which makes a comparison between them consistent.

\subsection{Results.}

\begin{figure*}
\includegraphics[width=0.32\textwidth]{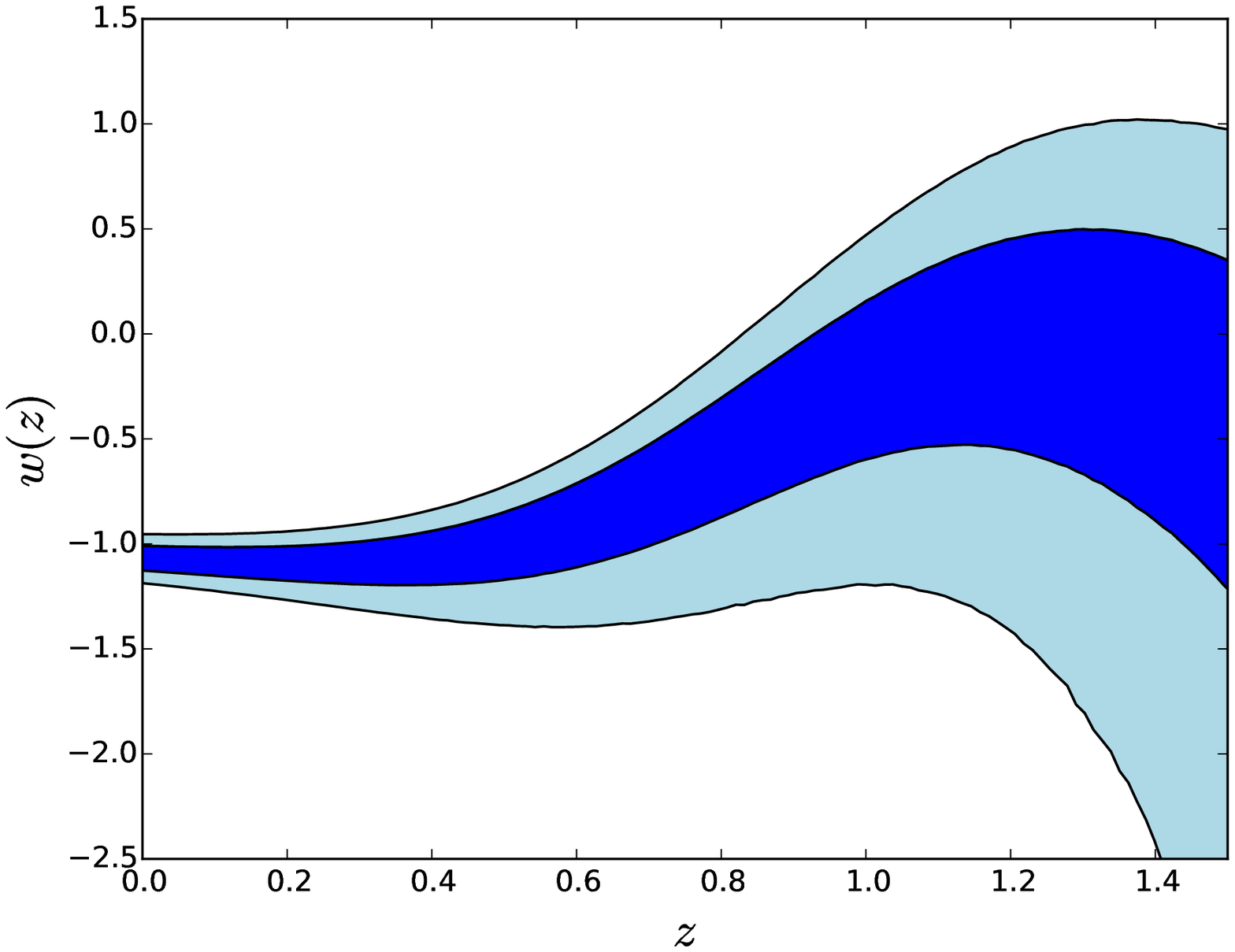}  \includegraphics[width=0.32\textwidth]{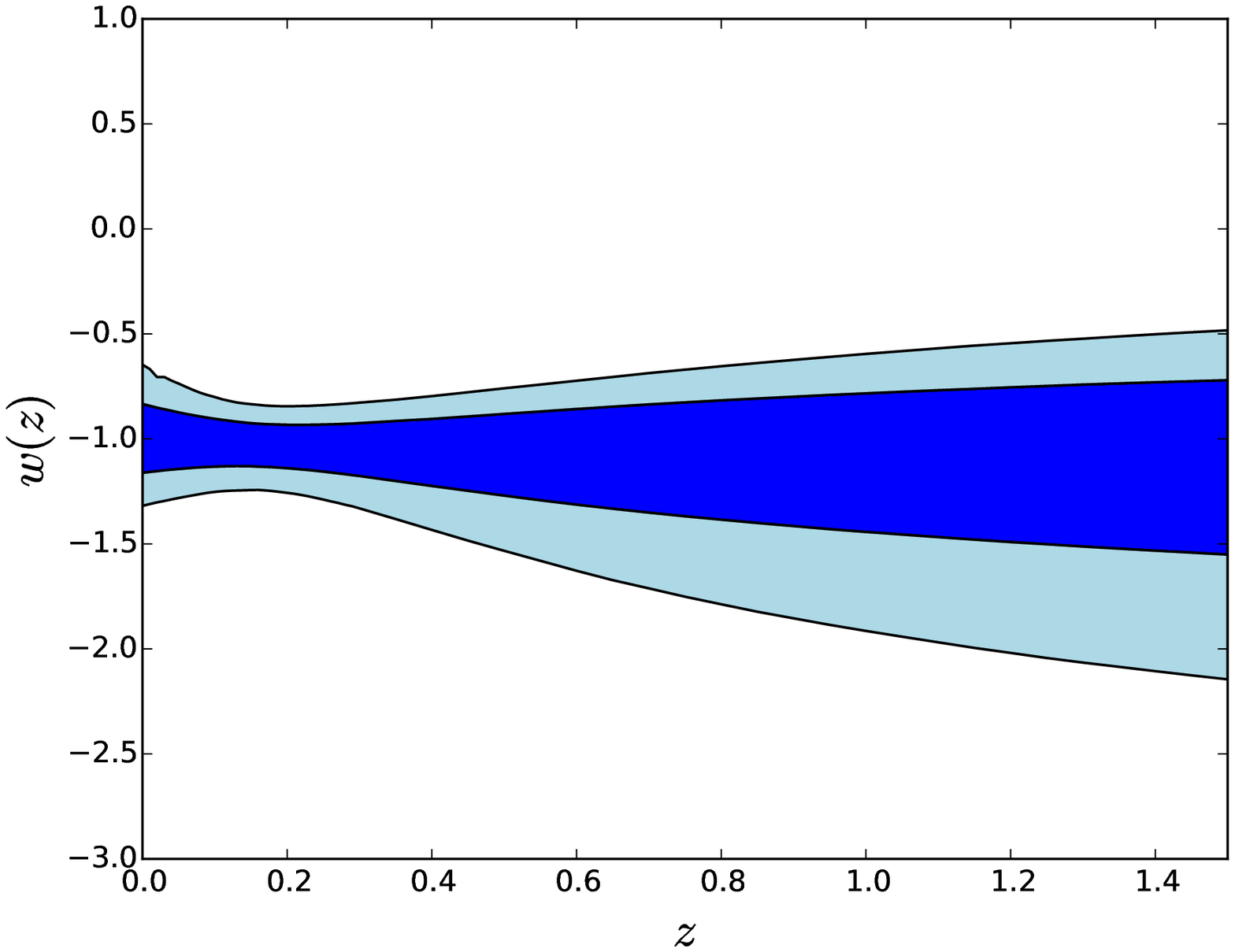} 
\includegraphics[width=0.32\textwidth]{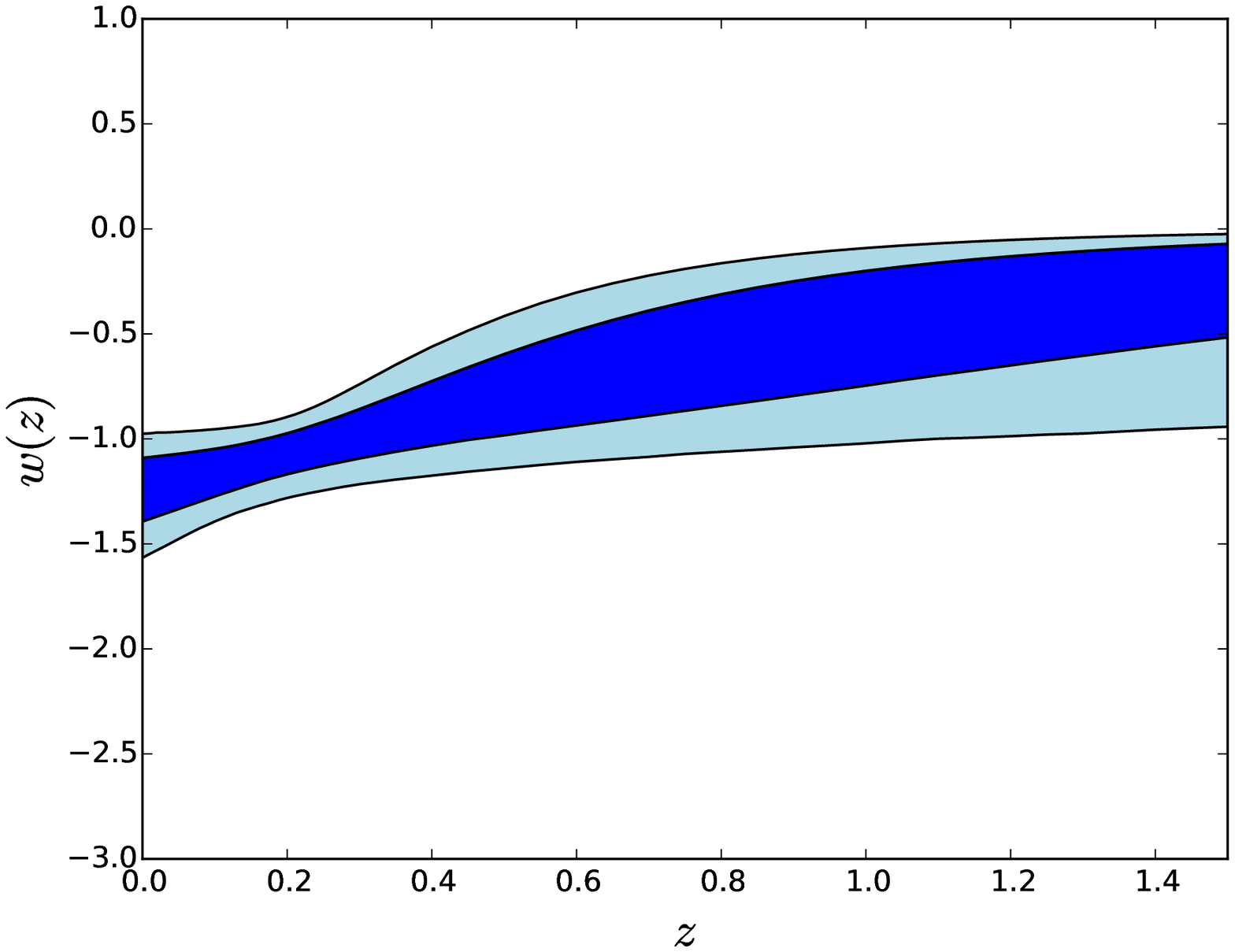}\\\includegraphics[width=0.32\textwidth]{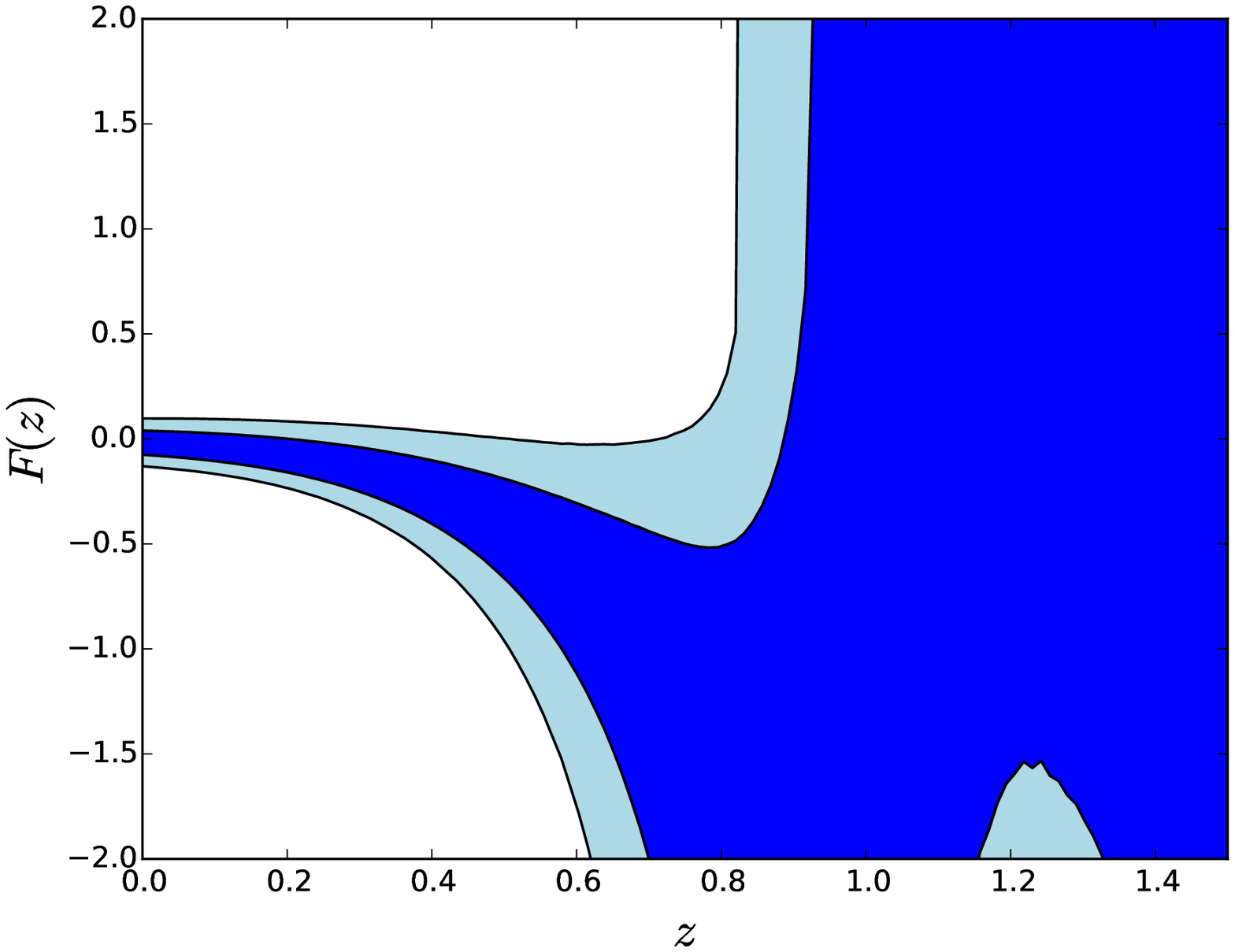}  
\includegraphics[width=0.32\textwidth]{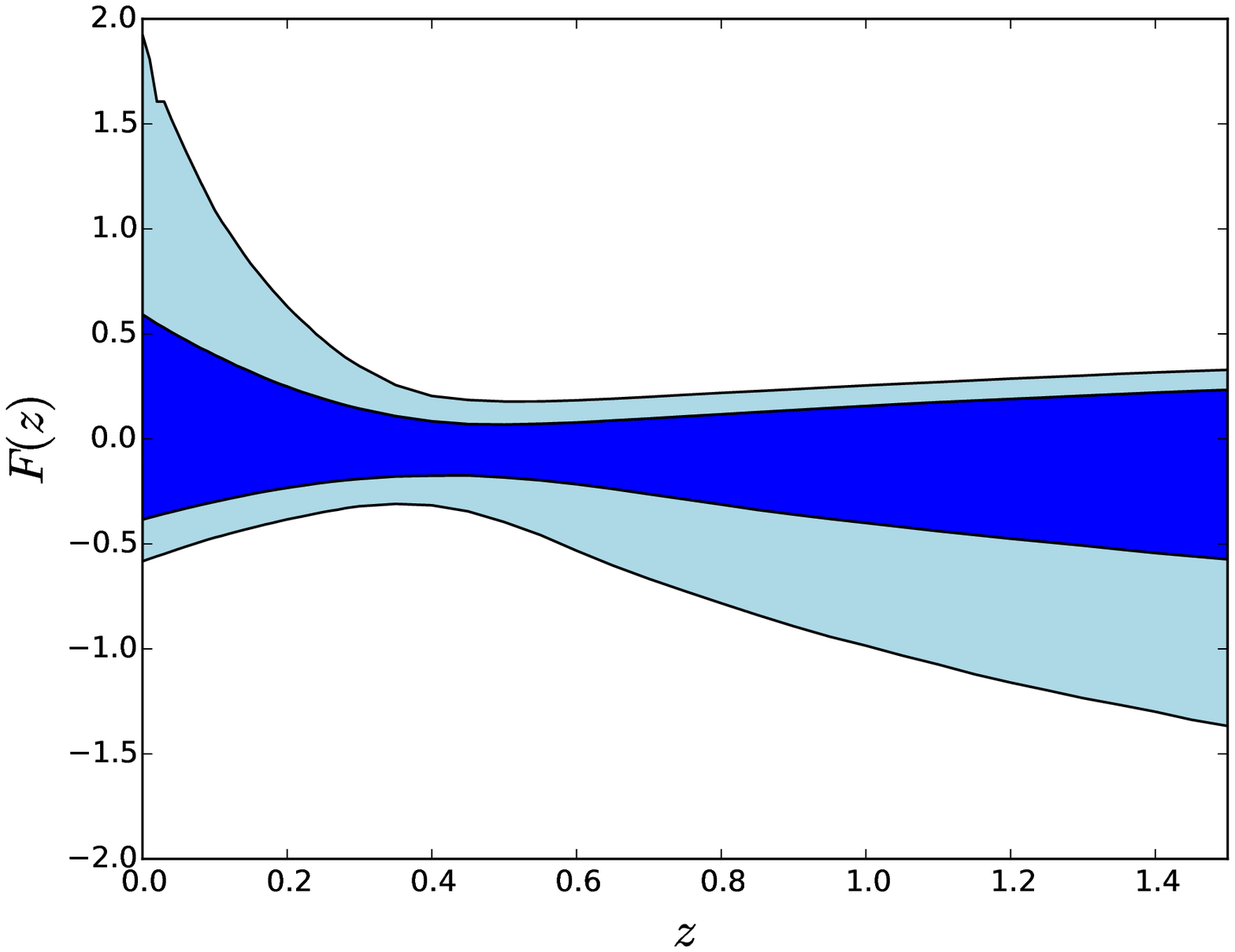}  \includegraphics[width=0.32\textwidth]{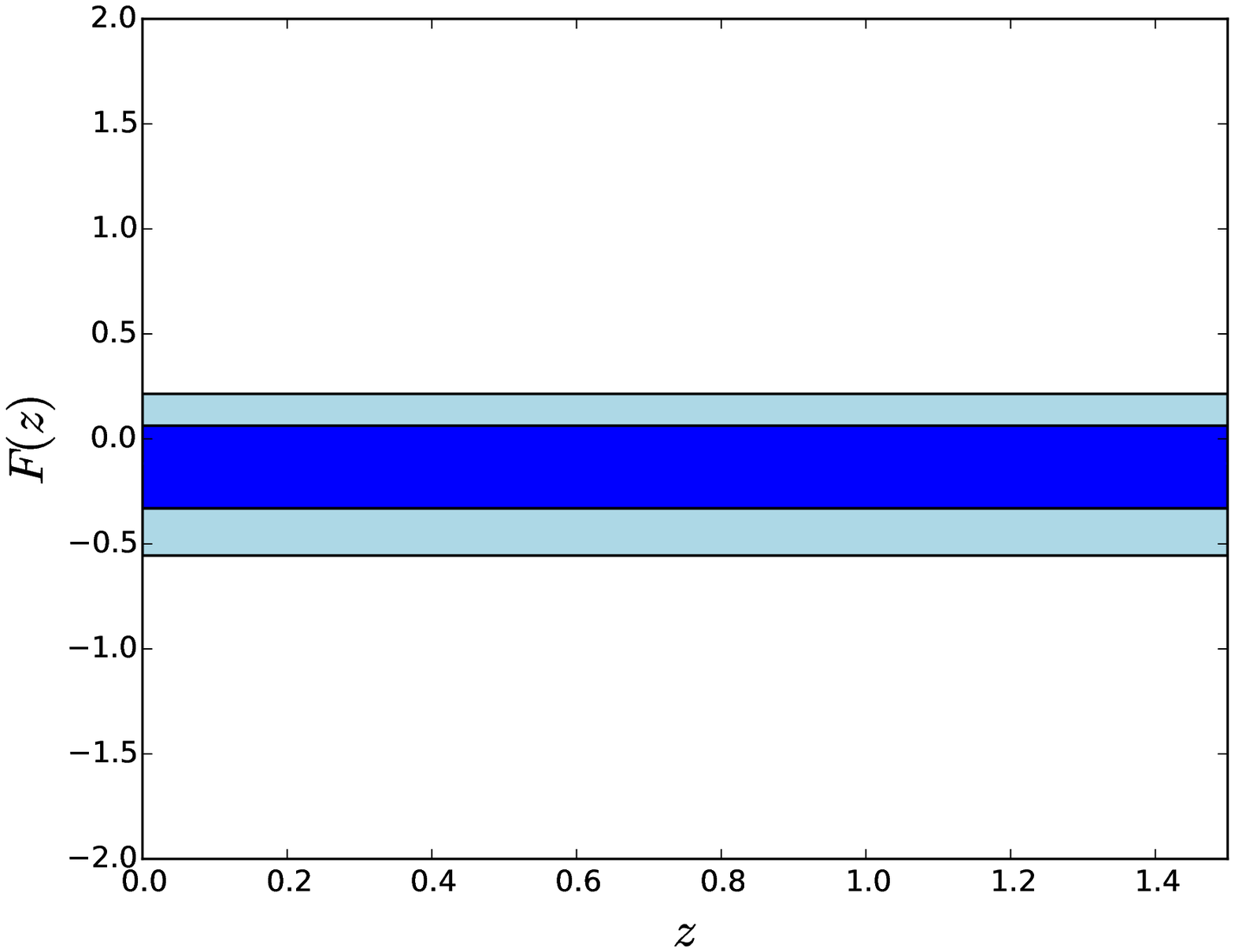}
\caption{$w(z)$ constraints, top and corresponding $F(z)$ constraints, bottom. The contours represent 68\% and 95\% confidence intervals. The left panels show the GP reconstruction. 
The middle panels depict the constraints the CPL parameterisation, while the right panels the constant F parameterisation.
A gaussian prior on $\Omega_m = 0.28 \pm 0.02$ is assumed for $w$, 
no prior is needed for $F$ (though for the parametric models, the contours narrow slightly with a prior on $\Omega_m$).}
\label{fig1}
\end{figure*}

\begin{figure*}
\includegraphics[width=0.32\textwidth]{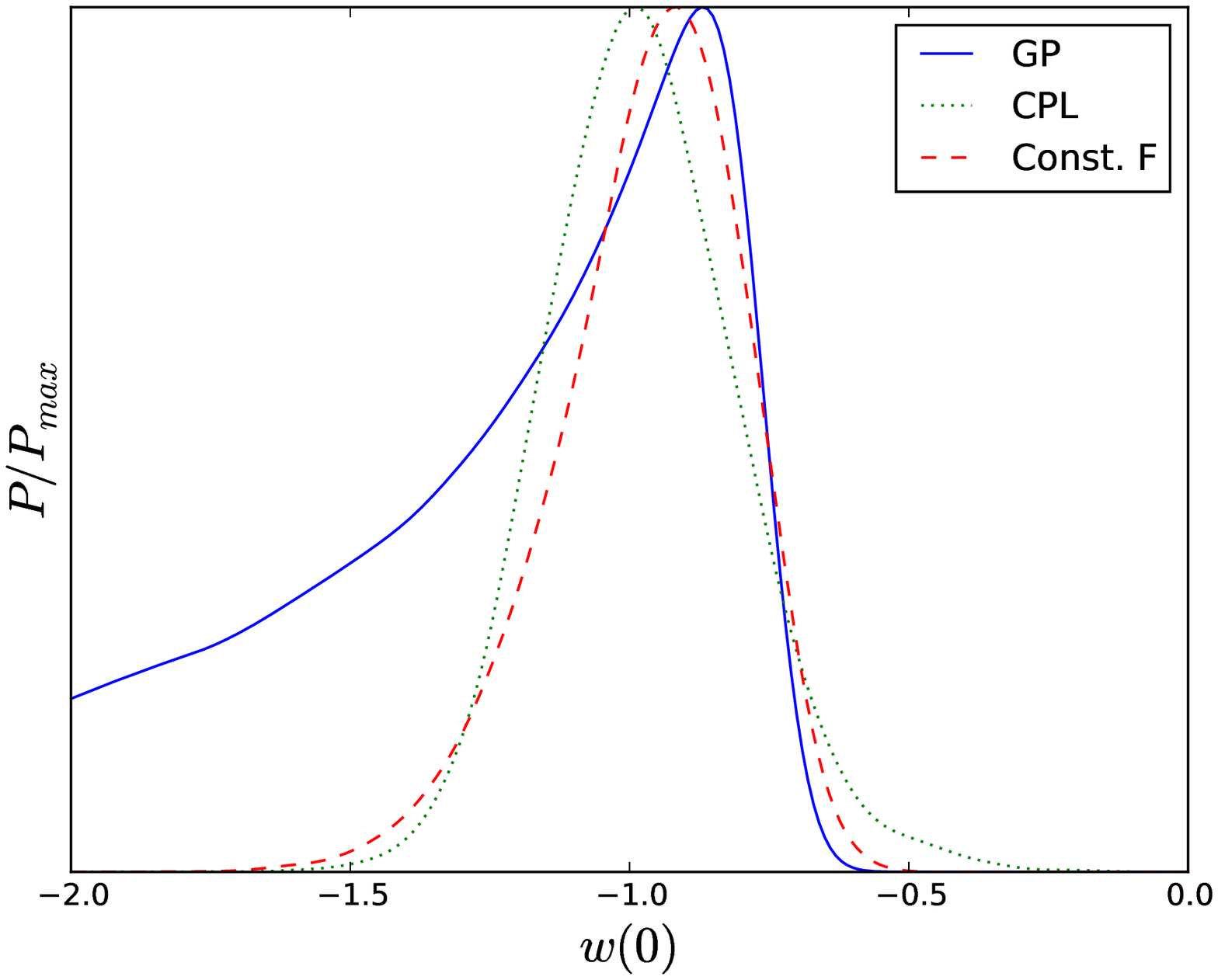}  \includegraphics[width=0.32\textwidth]{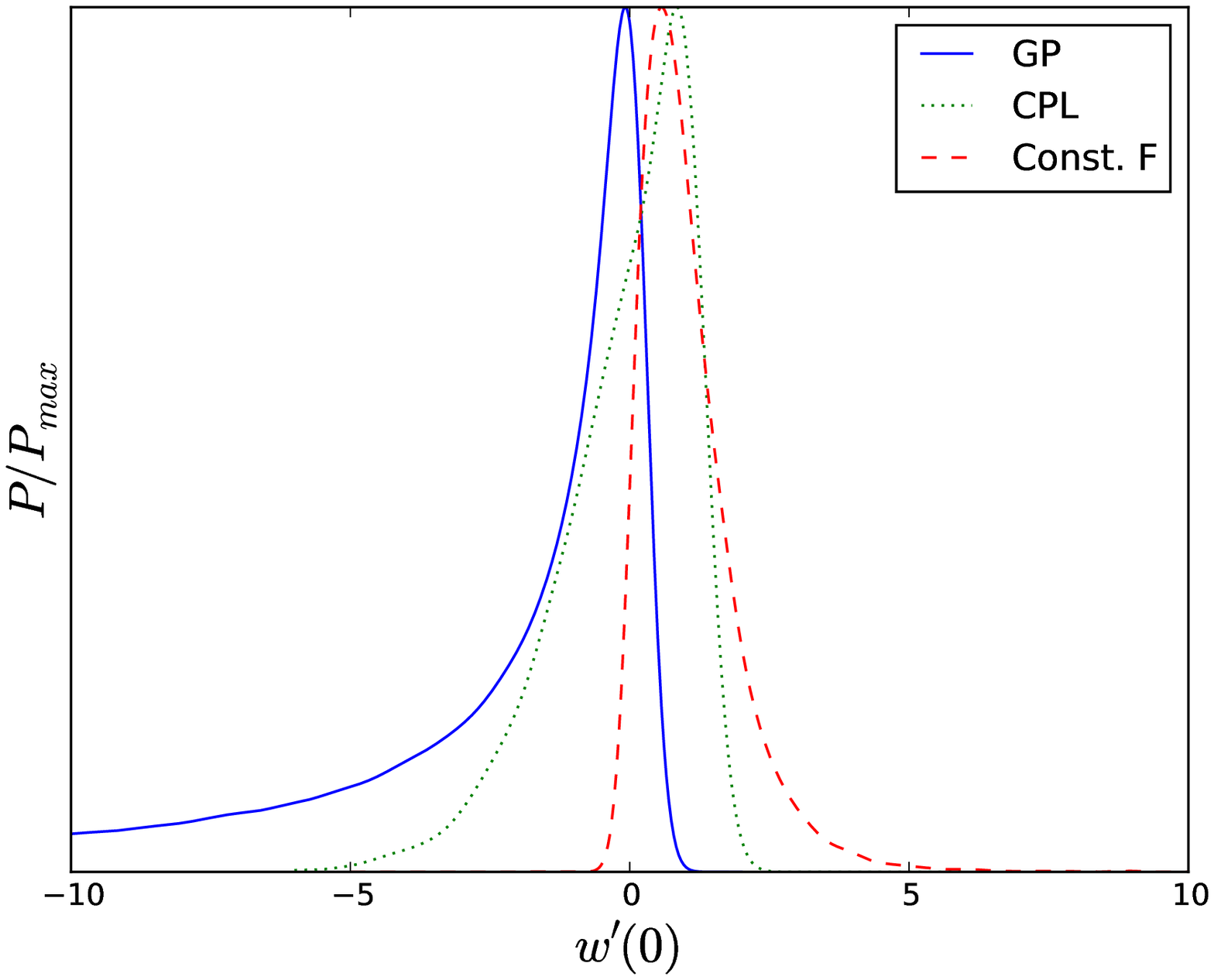} 
\includegraphics[width=0.32\textwidth]{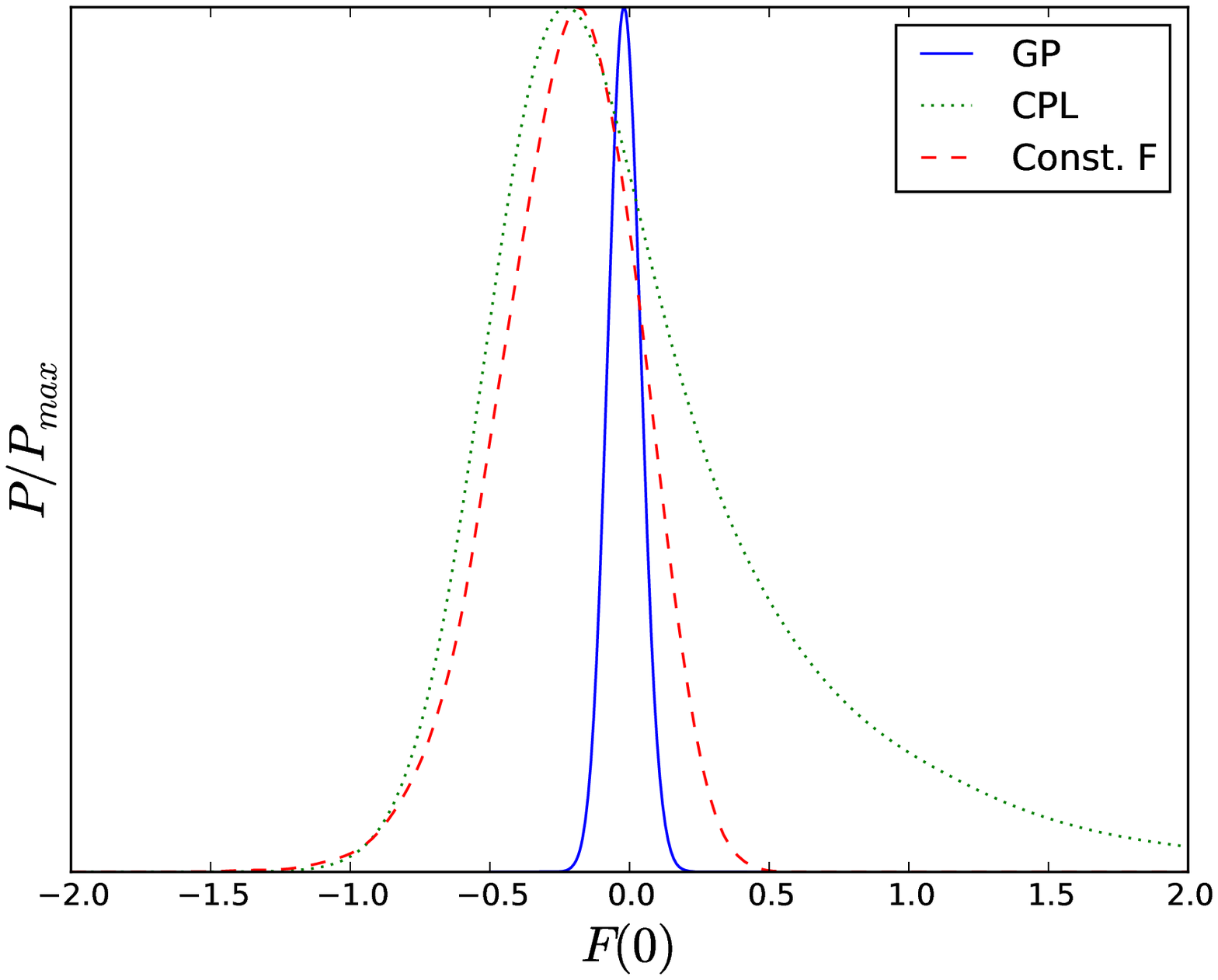}\\\includegraphics[width=0.32\textwidth]{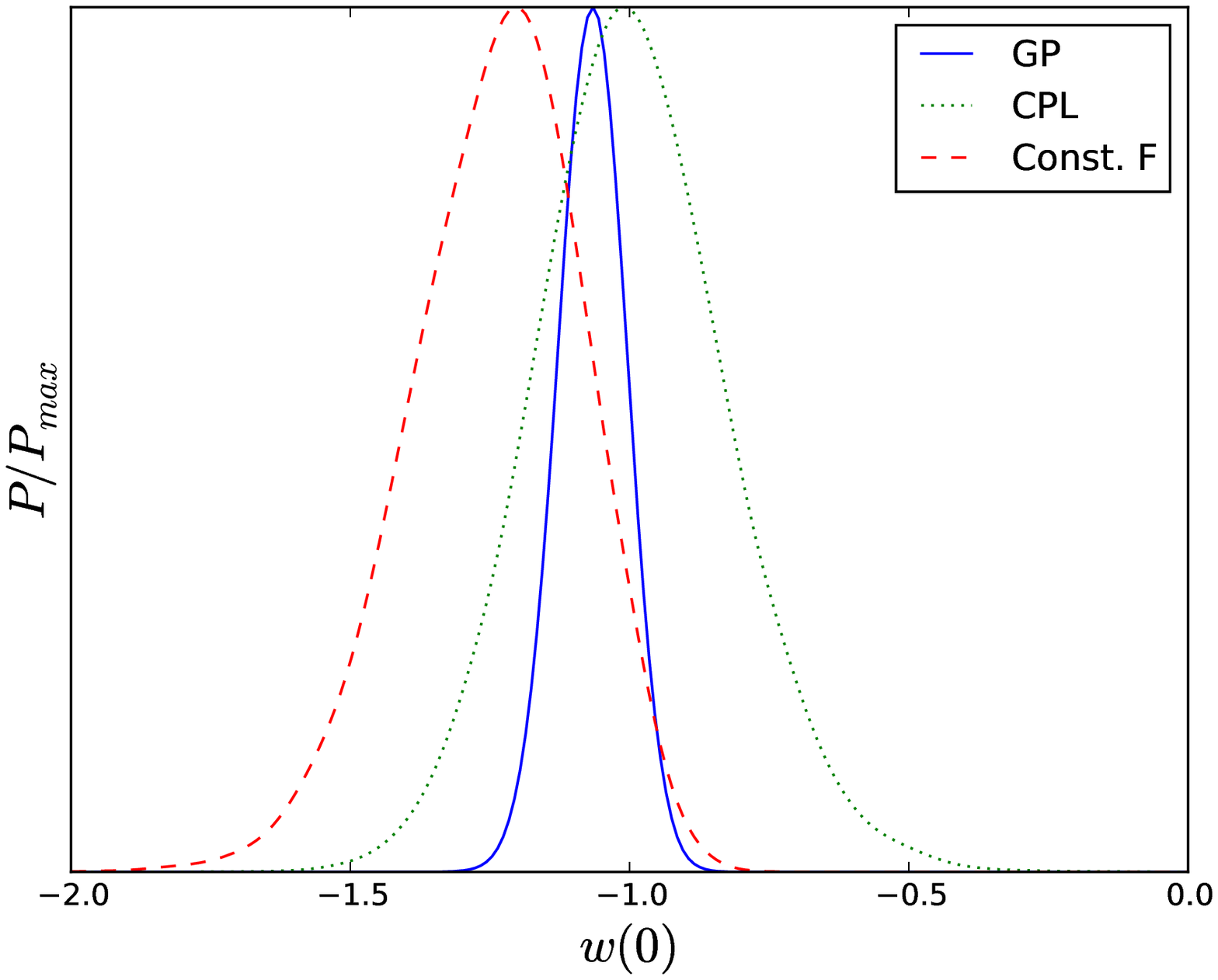}  
\includegraphics[width=0.32\textwidth]{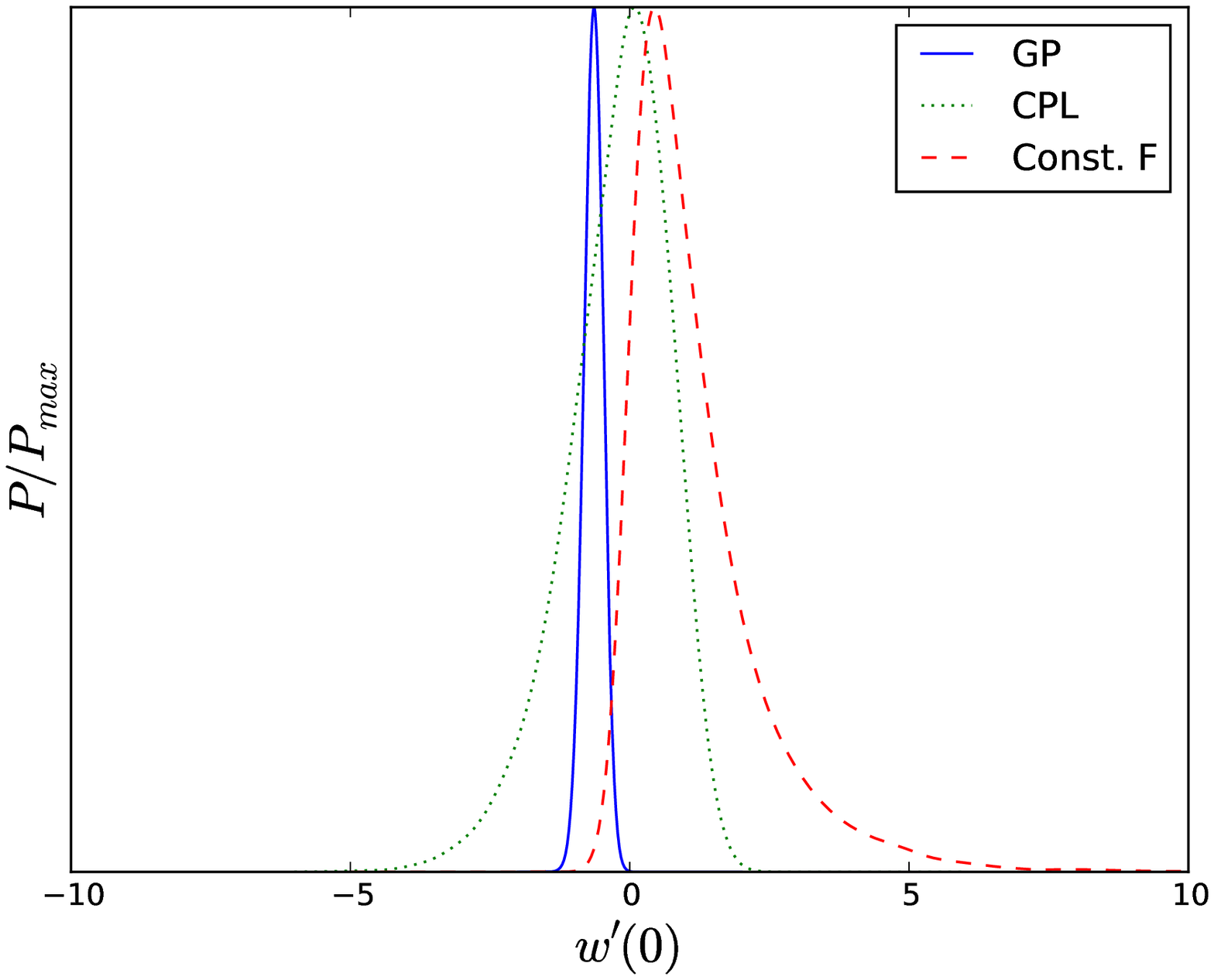}  \includegraphics[width=0.32\textwidth]{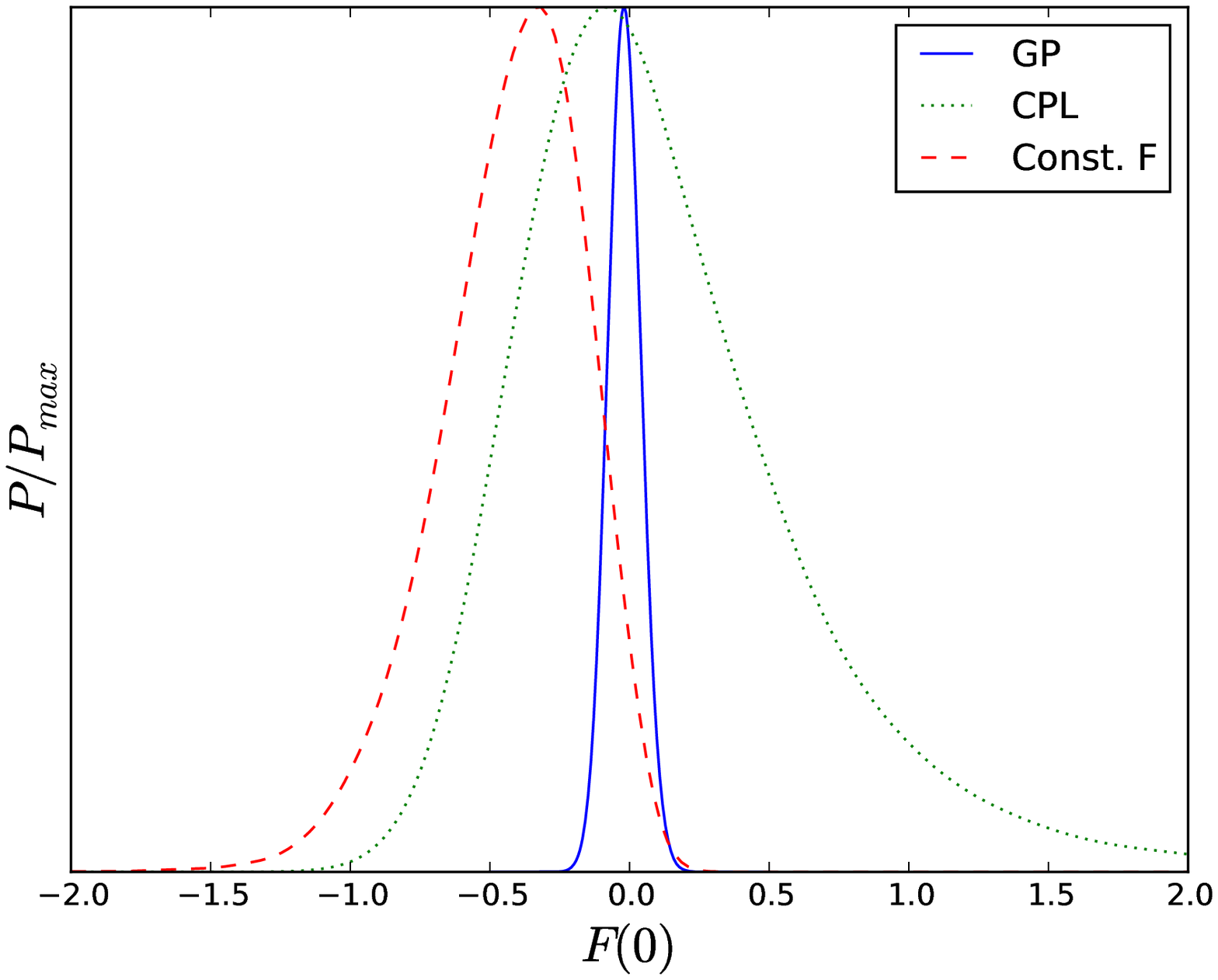}
\caption{Posterior for $w(0)$, $w'(0)$ and $F(0)$ found using Gaussian Processes, versus parametric approaches. 
The top row has a flat prior on $\Omega_m$, the bottom row has a Gaussian prior $\Omega_m = 0.28 \pm 0.02$. For all the parametric approaches the constraints are strongly 
dependent on the prior. For GPs, a prior is necessary for $w(0)$, $w'(0)$ but not $F$. For parametric approaches, constraints on $w(0)$ are reasonably consistent but achieve 
very different results for $w'(0)$ and $F(0)$, effectively giving no meaningful constraints on the evolution of the EOS at all. GPs on the other hand give the strongest 
constraints on all variables when a prior on $\Omega_m$ is included (it was shown that the errors on $w$ using this method are robust in~\cite{sc2013}). Strong constraints 
on $F$ are found using GPs independently of $\Omega_m$.}
\label{fig2}
\end{figure*}

In Fig.~\ref{fig1} we show constraints on $w(z)$,  and $F(z)$ for the different methods. The parametrised  models show broadly similar behaviour for the constraints 
on $w(z)$ and give similar errors compared to those found using GPs, at least for low to moderate $z$ (strong constraints at high $z$ only come from the choice of parameterisation). In this case, the good constraints with GPs are obtained because of the prior adopted on $\Omega_m$, where much weaker 
ones would be derived considering a broad flat prior on $\Omega_m$~-- this arises simply from~\eqref{djskbcskcb}. For $F(z)$ the situation is rather different, with very different constraints found for the different approaches. Constraints at all redshifts between parametrised models versus GPs are very different. With GPs $F(z)$ is well constrained at small redshifts $(z\lesssim0.5)$ and weakly constrained for redshifts 
above that. Also it is interesting to point out that for GPs the constraints on $F(z)$ are weakly dependent on the $H(z)$ data, where they  are slightly enlarged if we do not use this sample.
In redshift regions well sampled by data good constraints are derived, 
while redshift regions sparsely measured are not able to distinguish between different values for $F$, when combined with the fact that changes in $w$ affect $w$ at low $z$ most strongly. This is in stark contrast with the results from the parametric methods, 
which provide better results at
high redshifts due to the imposed shape of the function a priori.

In Fig.~\ref{fig2} we show the posteriors at $z=0$ for all the different cases. With no prior on $\Omega_m$ only $F$ found by GPs has strong constraints, while individual constraints on $w$ and $w'$ are significantly weaker. With a prior on $\Omega_m$, the constraints on $w(0)$, $w'(0)$ and $F(0)$ vary significantly. In particular, the parametric approaches favour both $w$ and $w'$ in different ways (CPL favours  $w$ nearer 0 and  $w'$ more negative, and the constant $F$ model pulls in the opposite directions), while GPs give tight constraints approaching the union of the parameterised models. 
 
It is important to stress one needs to be careful of not deriving biased results due to a given value for $\Omega_m$. For example, there is a tension for $w'(0)$ when using GPs and our (somewhat arbitrary) choice of prior on $\Omega_m$ which completely disappears when using {\it Planck} values \cite{planck} for $\Omega_m$ and $H_0$: $w'(0) = -0.03 \pm 0.23$ $(1\sigma)$. The $H(z)$ data prefer lower values of $H_0$ as it was shown in \cite{H0_bcs}. Therefore, GP reconstructed $F$ gives strong constraints favouring a $\Lambda$CDM model, with $F$  providing an unbiased, more powerful tool to look for deviations of the concordance model as a time-varying dark energy.

\subsection{Validating the method}

In order to analyse the robustness of our method, we performed 1000 simulations of SNe Ia data in a $\Lambda$CDM model 
with $\Omega_m=0.3$, $H_0=73.8$ km s$^{-1}$ Mpc$^{-1}$ and the same redshift distribution of the Union2.1 sample. For the sake of 
simplicity, we restricted ourselves only to SNe Ia data and we show the results for $F(z)$ with an emphasis on $F(0)$.

\begin{figure*}
\includegraphics[width=0.4\textwidth]{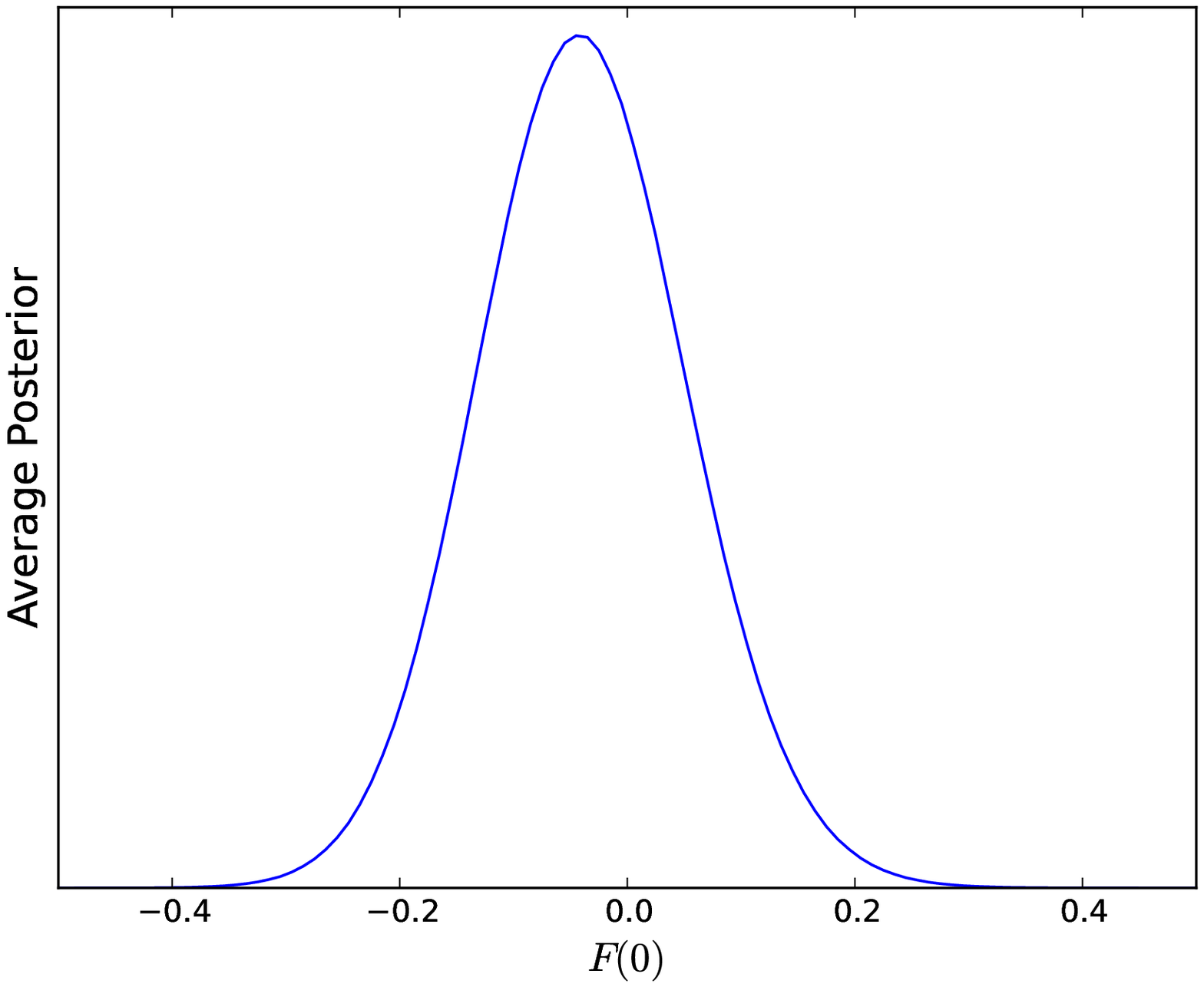}
\includegraphics[width=0.4\textwidth]{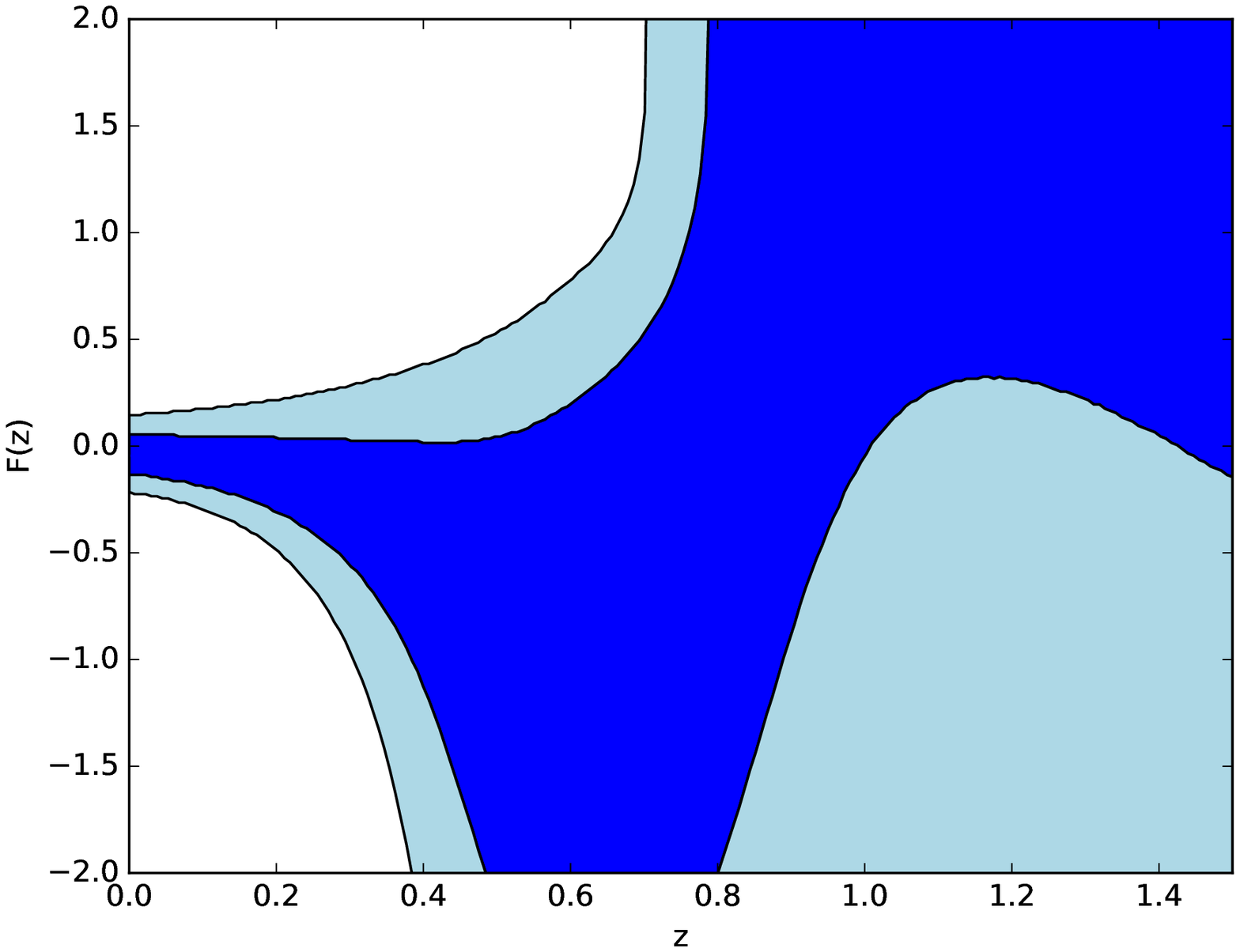}
\caption{Average posterior of one thousand simulations of SNe Ia data for a $\Lambda$CDM model with $\Omega_m=0.3$ and $H_0=73.8$ km s$^{-1}$ Mpc$^{-1}$. 
The same redshift distribution of the Union2.1 sample was used. In the left panel is shown the behaviour of $F$ at redshift zero, while in the right panel for the whole redshift
range. The size of errors is fully compatible with the observed data, apart from a small bias which can be neglected for the current sample size, but that should be better
characterized for bigger samples.}
\label{fig3}
\end{figure*}

Fig \ref{fig3} presents the average posterior for $F$ given the 1000 simulations, where the left panel has the result for $F(0)$ and the right panel the result for $F(z)$ for all redshifts.
As we can see, the width of the curve is 
consistent with the errors obtained, and there is a small bias much smaller than the errors. We also have checked that the results 
are not influenced by the fiducial cosmological model, since changing for XCDM or a time-varying equation of state for dark energy 
gave consistent results, as it is shown in Fig. 4. Bigger samples as the ones coming from the ongoing Dark Energy Survey will demand a thorough scrutiny of 
the bias, although it can be safely neglected for the current size of SNe Ia samples.

\begin{figure*}
\includegraphics[width=0.4\textwidth]{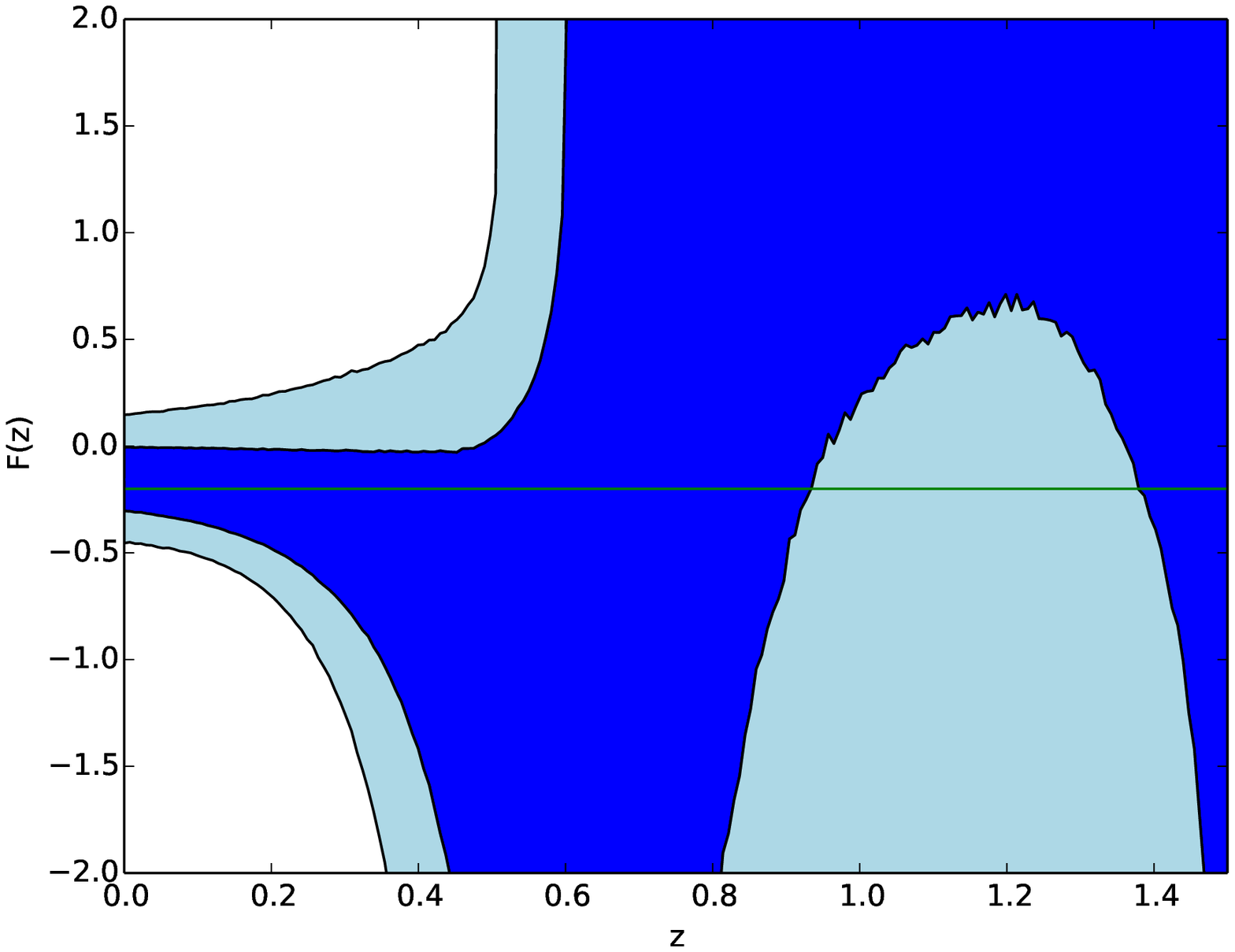}  
\includegraphics[width=0.4\textwidth]{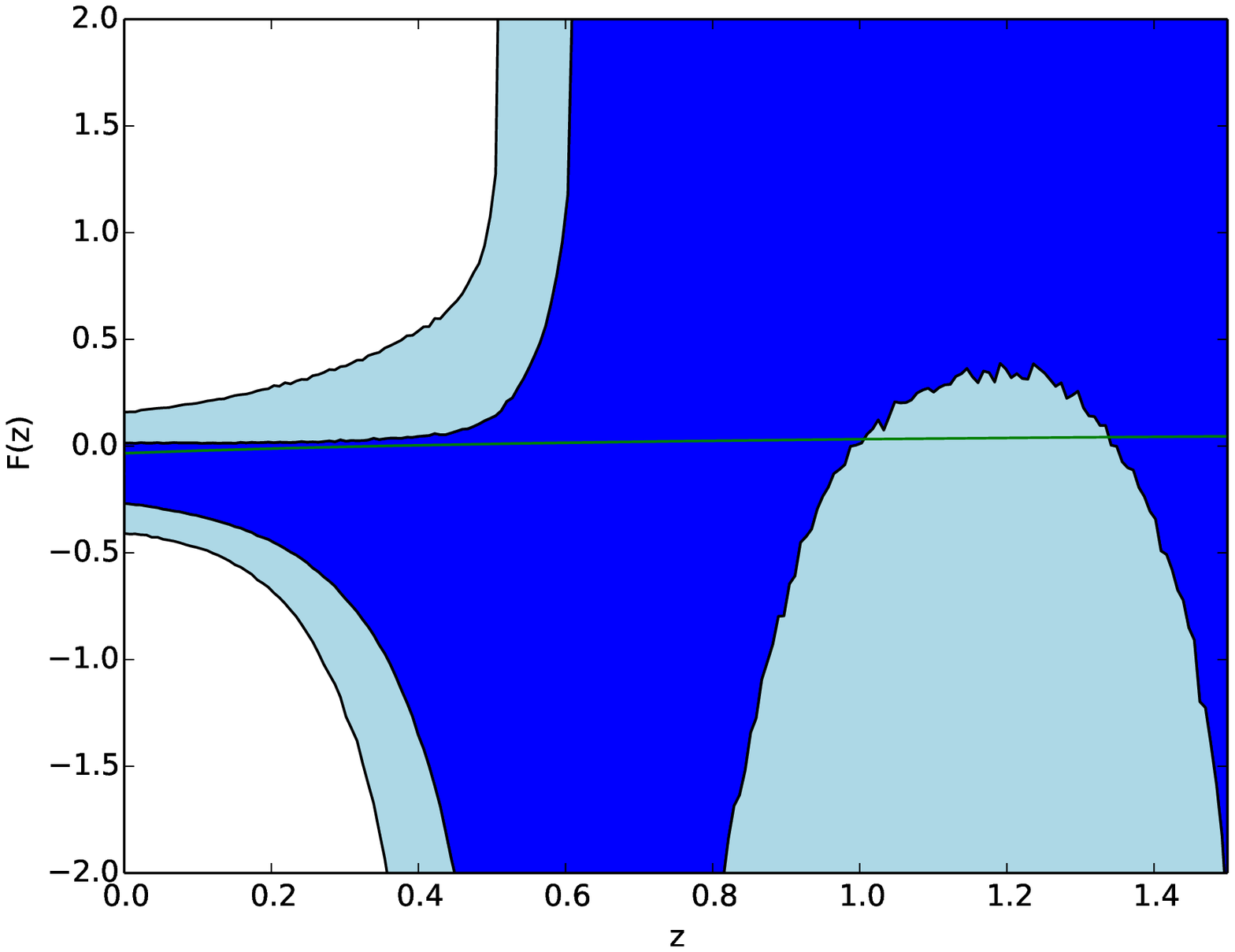}
\caption{Typical realization of SNe Ia data with different fiducial models. The left panel uses a fiducial model with constant $w = −1.1$ which gives
$F = −0.1$. The right panel uses a fiducial model with a time-varying equation of state with $w_0 = −1$
and $w_a = 0.1$. The solid line shows the value of the expected model.}
\label{fig4}
\end{figure*}

\section{Conclusions}

We have shown that for a flat model there exists a combination of $w$ and its first derivative $w'$ which can be found purely from Hubble normalised distance data independently of $\Omega_m$. This provides 
a route by which we can place constraints on the time evolution of $w$ without requiring a prior on $\Omega_m$. At $z=0$ this corresponds to roughly the lines in the $w_0-w_a$ 
plane where $w_a/3w_0+1+w_0=$const. We have shown that using GPs strong constraints can be placed on $F(0)$ using current data, and with no assumptions on $\Omega_m$ or $w(z)$. This is to be seen in stark contrast to parametric models which have significantly weaker constraints, but an order of magnitude. Furthermore, the constraints on $F$ are similarly much stronger than constraints on either $w$ or $w'$ at the origin. 
Consequently, using $F$ as a complimentary function to $w(z)$ might be a more useful way to characterize the time evolution of $w$ than is currently done.
 
Of course, if one wants $w(z)$ explicitly the degeneracy with $\Omega_m$ is unavoidable. In this sense, have we really dodged the degeneracy? 
We have presented a new perspective on the question: since $w(z)$ is usually a placeholder for whatever is `not-$\Lambda$', it may well be 
a sub-optimal measure of `not-$\Lambda$'. A set of observables such as $[F(z),F'(z),F''(z),\ldots]$, formed from~\eqref{eqF} rephrase 
the $\Lambda$ question as: do $[F(z),F'(z),F''(z),\ldots]_{z=0}=0$? but now independently of $\Omega_m$.

~\\
\paragraph*{Acknowledgements.}
The authors thank M. Seikel for her participation during early stages of this work. VCB thanks Michel Aguena da Silva for his help with MCMC. 
VCB was supported by CNPq-Brazil, with a fellowship within the program Science without Borders, FAPESP (grant number 2014/21098-1) and CAPES. CC is funded by the National Research Foundation (South Africa).

\appendix

\section{Gaussian processes}

A Gaussian process is defined by a collection of random variables, any finite number of
which have (consistent) joint gaussian distributions \cite{rasmussen}. Basically speaking,
while a gaussian variable is a distribution over random variables, a gaussian process is a
distributions over random functions, characterized by a mean and a covariance function.

In order to reconstruct a given function, we assume that this function is a realization of a GP
with a given mean and covariance function. Assume we have $N$ observational data points $\bm y=(y_1, \dots, y_N)$
of a function $f(z)$ at redshifts $\bm Z=(z_1, \dots, z_N)$. The
errors of the observations are given in the covariance matrix
$C$. Additionally we have $\tilde{N}$ observations of the first
derivative $f'(z)$, given by $\tilde{\bm Z}=(\tilde{z_1}, \dots,
\tilde{z}_{\tilde{N}})$, $\tilde{\bm y}'=(\tilde{y}'_1, \dots,
\tilde{y}'_{\tilde{N}})$ and $\tilde{C}$. Thus, the goal is to reconstruct
the function which underlies the data at redshifts $\bm Z^*=(z^*_1,
\dots, z^*_N)$. We denote these function points as $\bm
f^*=(f(z^*_1), \dots, f(z^*_N))$ and the $n$th derivative of the
function at these redshifts as $\bm f^{*(n)}$.

Generally, we incorporate the desired properties of the function under study in the covariance function.
In this work we adopt the Mat\'ern($\nu=9/2$) covariance function (see \cite{sc2013} for a discussion on the impact of the
covariance function)

\begin{eqnarray}
k(z_i,z_j) &=& \sigma_f^2
  \exp\left[-\frac{3\,|z_i-z_j|}{\ell}\right]  \\
  && \times \left(1 +
  \frac{3\,|z_i-z_j|}{\ell} + \frac{27(z_i-z_j)^2}{7\ell^2}\right. \nonumber\\
&&\qquad \left. {}+ \frac{18\,|z_i-z_j|^3}{7\ell^3} +
  \frac{27(z_i-z_j)^4}{35\ell^4} \right)\nonumber \, ,
\end{eqnarray}
where we have two hyperparameters $\sigma_f$ and
$\ell$ which describe how the function changes in the $y$ and $x$ axis, respectively.

\begin{widetext}

The probability distribution for the points we want to reconstruct is gaussian with the mean and covariance given by

\begin{equation}
\overline{{\bm f}^{*(n)}} = 
\begin{pmatrix}
K^{(n,0)}(\bm Z^*,\bm Z) &  K^{(n,1)}(\bm Z^*,\tilde{\bm Z})
\end{pmatrix}
\begin{bmatrix}
K(\bm Z,\bm Z) + C            & K^{(0,1)}(\bm Z,\tilde{\bm Z}) \\
K^{(1,0)}(\tilde{\bm Z},\bm Z)  & K^{(1,1)}(\tilde{\bm Z}, \tilde{\bm Z}) + \tilde{C}
\end{bmatrix}^{-1}
\begin{pmatrix}
{\bm y}\\
\tilde{\bm y}'
\end{pmatrix}
\end{equation}
and
\begin{equation}
\text{cov}\left({\bm f}^{*(n)}\right) = K^{(n,n)}(\bm Z^*,\bm Z^*) - 
\begin{pmatrix}
K^{(n,0)}(\bm Z^*,\bm Z) &  K^{(n,1)}(\bm Z^*,\tilde{\bm Z})
\end{pmatrix}
\begin{bmatrix}
K(\bm Z,\bm Z) + C            & K^{(0,1)}(\bm Z,\tilde{\bm Z}) \\
K^{(1,0)}(\tilde{\bm Z},\bm Z)  & K^{(1,1)}(\tilde{\bm Z}, \tilde{\bm Z}) + \tilde{C}
\end{bmatrix}^{-1}
\begin{pmatrix}
K^{(0,n)}(\bm Z,\bm Z^*) \\
K^{(1,n)}(\tilde{\bm Z},\bm Z^*)
\end{pmatrix} \, .
\end{equation}
Here, the covariances are written as matrices $K(\bm X,\tilde{\bm X})$
with $\{K(\bm X,\tilde{\bm X})\}_{ij} = k(x_i,
\tilde{x}_j)$. $k^{(n,m)}$ denotes the $n$th derivative of $k$ with
respect to the first argument and the $m$th derivative with respect to
the second argument.

The hyperparameters are determined by maximizing the likelihood
\begin{eqnarray}\label{log-marginal-p}
\ln \mathcal{L} &=& \ln p({\bm y}|\bm Z,\tilde{\bm Z},\sigma_f,\ell) \\
&=& -\frac{1}{2}
\begin{pmatrix}
{\bm y}^T & \tilde{\bm y}'^T
\end{pmatrix}
\begin{bmatrix}
K(\bm Z,\bm Z) + C            & K^{(0,1)}(\bm Z,\tilde{\bm Z}) \\
K^{(1,0)}(\tilde{\bm Z},\bm Z)  & K^{(1,1)}(\tilde{\bm Z}, \tilde{\bm Z}) + \tilde{C}
\end{bmatrix}^{-1}
\begin{pmatrix}
{\bm y}\\
\tilde{\bm y}'
\end{pmatrix}
- \frac{1}{2}\ln
  \left|
\begin{matrix}
K(\bm Z,\bm Z) + C            & K^{(0,1)}(\bm Z,\tilde{\bm Z}) \\
K^{(1,0)}(\tilde{\bm Z},\bm Z)  & K^{(1,1)}(\tilde{\bm Z}, \tilde{\bm Z}) + \tilde{C}
\end{matrix}
\right|
  - \frac{{\tilde{N}}+N}{2}\ln 2\pi \, . 
\nonumber
\end{eqnarray}

While the best approach would be to marginalize over the hyperparameters, it was shown in \cite{sc2013} that for a sample of the same size considered in this work, 
maximization provides basically indistinguishable results at low redshifts, which is the region where good constraints were derived.

We use GaPP (Gaussian Processes in Python) \cite{gp2} to derive the GP results.
The code is free to use and was applied in many situations \cite{gp_examples}.

\end{widetext}


\begin{thebibliography}{99}
\bibitem{foot} Note that $F=0$ does not uniquely imply $\Lambda$CDM because it is degenerate with 
$w(z)=w_0/[(1+w_0)(1+z)^3-w_0]$, but this does not have a clear physical motivation.
\end{thebibliography}
\end{document}